%Paper: hep-th/9506012
%From: jerome@theory.caltech.edu (Jerome Gauntlett)
%Date: Thu, 1 Jun 95 18:52:01 PDT
%Date (revised): Thu, 1 Jun 95 19:13:57 PDT

%%%%%%%%%%%This is a Plain TEX file that uses vanilla.sty%%%%%%%%%%%%%%%%%%%%

%UPDATED BY RICHARD WITH EFFECT FROM: 27 OCTOBER 1994

%------------------------------------------------------------------------
\input vanilla.sty
\headline={\ifnum\pageno=1\firstheadline\else
\ifodd\pageno\rightheadline \else\leftheadline\fi\fi}
\def\firstheadline{\hfil}
\def\rightheadline{\hfil}
\def\leftheadline{\hfil}
	\footline={\ifnum\pageno=1\firstfootline\else\otherfootline\fi}
\def\firstfootline{\rm\hss\folio\hss}
\def\otherfootline{\hfil}

 1
 1
 1

\font\tenbf=cmbx10
\font\tenrm=cmr10

\font\eightrm=cmr8
\font\eightit=cmti8

\parindent=1.2pc
\magnification=\magstep1
\hsize=6.0truein
\rightline{CALT-68-1997}
\rightline{hep-th/9506012}
\medskip
\vsize=8.6truein
\nopagenumbers
%\rightline{CALT-68-1997}

\centerline{\tenbf $S$-Duality and $H$-Monopoles\footnote"$^a$"{
\eightrm\baselineskip=10pt This is based on work done in collaboration with
J. Harvey; see Ref. (1) for further details and references.}}
%\baselineskip=18pt
%\centerline{\tenbf GUIDELINES FOR TYPESETTING A CAMERA-READY}
\baselineskip=13pt
%\centerline{\tenbf MANUSCRIPT BY COMPUTER}
%\centerline{\eightss (For subsequent 20\% photoreduction
%to 17.8$\times$11.9 cm text area)\footnote"$^*$"{\eightrm\baselineskip=10pt
%The \TeX\ source file for this document may be used as a template
%for your article, and can be requested by e-mailing {\eightss
%worldscp\@singnet.com.sg}.}}

\centerline{\eightrm Jerome P. Gauntlett}
\baselineskip=12pt
\centerline{\eightit California Institute of Technology, Mail Code 452-48}
\baselineskip=10pt
\centerline{\eightit Pasadena, CA, 91125, U.S.A.}
\centerline{\eightrm E-mail: jerome\@theory.caltech.edu}
%\vglue0.2cm
%\centerline{\eightrm and}
%\vglue0.2cm
%\centerline{\eightrm SECOND AUTHOR'S NAME}
%\baselineskip=12pt
%\centerline{\eightit Group, Company, Address, City,
%State ZIP/Zone, Country}

\vglue0.6cm
%\centerline{\eightrm ABSTRACT}
%\vglue0.2cm
%{\rightskip=3pc
 %\leftskip=3pc
 %\eightrm\baselineskip=10pt\noindent
%The spectrum of $H$-monopoles and their relationship to the
%conjecture that the four dimensional heterotic string theory
%compactified on a torus is $S$-dual is reviewed and updated.
%It is based on work done in collaboration with J. Harvey.
%\vglue0.6cm}

\tenrm\baselineskip=13pt
%\leftline{\tenbf 1. General Appearance}
%\vglue0.4cm

It has been conjectured that
the four dimensional
heterotic string theory compactified on a torus
is $SL(2,Z)$ invariant or ``$S$-dual". The duality
group, which acts on the axion, dilaton and the gauge fields,
includes strong-weak coupling and electric-magnetic
duality as a special case. A consequence of the conjecture is that
the spectrum of
BPS states, states that saturate a Bogomol'nyi bound and hence form
short representations of the $N=4$ supersymmetry algebra, should be $S$-dual.
It was pointed out by Sen$^2$ that this leads to specific
predictions about the spectrum of magnetic monopoles some of which
may be testable at weak coupling.

The four dimensional theory has an
infinite tower of electrically charged BPS states in which
the right movers are in their ground state.
These states
obey the constraint $N_L-1 =(p^2_R-p^2_L)/2$
and obey the mass shell
condition
$
M=p^2_R/2
$
where $(p_L,p_R)\in \Gamma_{22,6}$ with $\Gamma_{22,6}$
being an even self-dual lattice specifying the compactification (we
consider generic points in the Narain moduli space ${\cal M}_N$
where the gauge group is
$U(1)^{28}$).
Acting with $S$-duality on these states we predict an infinite tower of
magnetic monopoles and dyons in the spectrum. For $N_L=0$ the
predicted spectrum is the same as in $N=4$ super Yang-Mills
theory and has been verified in the field theory limit.
For $N_L>1$ there is no region in ${\cal M}_N$ where the states
are light
and so we don't expect to find evidence for them in the field theory
limit. For $N_L=1$
the states can be light and so in particular
we expect
to find the magnetically charged $S$-duals in the field theory limit
using semi-classical
techniques.
These are the $H$-monopoles,
states that are magnetically charged with respect to $U(1)$'s coming
from the dimensional reduction of the antisymmetric tensor field.
Thus, it would seem that the spectrum of $H$-monopoles provides
an important
window into testing the $S$-duality conjecture in string theory
over and above its validity in $N=4$ super-Yang-Mills theory.

More precisely, the $S$-duality conjecture predicts that there are
24 short $N=4$
multiplets
of $H$-monopoles corresponding to the 24 ways of
choosing $N_L=1$.
The classical $H$-monopole solutions are based on instantons on $R^3\times S^1$
via the Bianchi identity $dH=\alpha' Tr F\wedge F$.
Since we are interested in the case
when all the non-abelian fields are broken down to $U(1)$'s, we construct
the solutions by picking an $SU(2)$ instanton with
non-trivial holonomy around the $S^1$. The moduli
space of $H$-monopoles, $M$, is then the moduli space of such instantons.
Incorporating the fermion zero modes, one expects that the
low-lying spectrum of $H$-monopoles is
obtained by studying
$N=4$ supersymmetric quantum mechanics on $M$.

Translation invariance implies that the moduli space is of the form
$M=R^3\times S^1\times \tilde M$. The four collective coordinates parametrising
$\tilde M$ come from the scale size of the instanton and from broken
spatial rotations.
{}From the supersymmetry we deduce that $\tilde M$ must be hyperK\"ahler and
from
rotation invariance that it should have $SO(3)$ isometry. Finally $\tilde M$
should have a $Z_2$ orbifold singularity corresponding to
instantons with zero scale size. From the analysis of Ref. (3) this restricts
$\tilde M$ to be one of three cases\footnote"$^b$"{This corrects a
statement made in Ref. (1).}: 1) $R^4/Z_2$; 2) Taub-Nut space; 3)
A one parameter family of metrics with additional curvature singularites (these
singularities seem difficult to interpret and can possibly be used to rule
out this case).

Quantising $N=4$ supersymmetric quantum mechanics on {\it any} of these
candidates does not seem to lead to 24 short multiplets;
what is required is the
existence of 24 normalisable harmonic forms on $\tilde M$.
In Ref. (1) we argued that
if one treats collective coordinates by doing string theory instead of
quantum mechanics on the moduli space and if the moduli space is $R^4/Z_2$
then we can see evidence for 24 states using orbifold techniques. However,
there are severe infrared problems to deal with which makes the outcome
of this proposal
inconclusive.

In conclusion, despite naive expectations,
it seems difficult to find convincing evidence
for (or against) the $S$-duality conjecture
by studying the spectrum of $H$-monopoles.
We close with some comments.
In our construction of the $H$-monopoles we worked near a point
in ${\cal M}_N$ with enhanced gauge symmetry. At a generic point in
${\cal M}_N$
one might
expect from scaling arguments that the generic
$H$-monopoles always shrink to zero scale size.
In this case a field theory analysis would not
be sufficient to determine the spectrum. Note that we
have not {\it rigorously} ruled this option out in our
construction, although we think it is unlikely.
It is also perhaps worth pointing out that the origin of the 24 states
is intimately connected to the gauge group of the ten dimensional heterotic
string being
$E_8\times E_8$ or $SO(32)$; maybe these groups should somehow
enter into the analysis.
If the $S$-duality conjecture is correct then it is clear that there
is more to be understood about $H$-monopoles.

\vglue0.6cm
\leftline{\tenbf References}
\vglue0.4cm

\itemitem{1.} J. P. Gauntlett and J. A. Harvey, hep-th/9407111.

\itemitem{2.} A. Sen, Int. J. Mod. Phys. {\bf A9} (1994) 3707.

\itemitem{3.} M. F. Atiyah and N. J. Hitchin,
{\it The Geometry and Dynamics of Magnetic Monopoles
\/} (Princeton University Press, Princeton, 1988), p. 70.

\bye